\documentclass[sigconf,nonacm]{acmart}

\usepackage{algorithmic}
\usepackage{graphicx}
\usepackage{textcomp}
\usepackage{xcolor}
\def\BibTeX{{\rm B\kern-.05em{\sc i\kern-.025em b}\kern-.08em
    T\kern-.1667em\lower.7ex\hbox{E}\kern-.125emX}}
    
\usepackage{subfigure}
\usepackage{comment}
\usepackage{booktabs}
\usepackage{balance}
\usepackage{pgfplots}
\usepackage{multirow}
\usepackage{url}
\usepackage{array}
\usepackage{colortbl}
\usepackage{adjustbox}
\usepackage{enumitem}
\usepackage[lined,linesnumbered,boxed,vlined,ruled]{algorithm2e}
\usepackage{caption}
\SetKwRepeat{Do}{do}{while}
\pgfplotsset{compat=1.16}
\definecolor{pastelred}{rgb}{1.0, 0.41, 0.38}

\newcolumntype{L}[1]{>{\raggedright\arraybackslash}p{#1}}
\newcolumntype{C}[1]{>{\centering\arraybackslash}p{#1}}
\newcolumntype{R}[1]{>{\raggedleft\arraybackslash}p{#1}}

\newcommand{\stitle}[1]{\vspace{0.5ex}\noindent{\bf #1}}

\newcommand{\system}{\textsc{Rpt}\xspace}
\newcommand{\pbi}{\textsc{Pbi}\xspace}

\begin{document}

\title{\underline{\system}: Effective and Efficient \underline{R}etrieval of \underline{P}rogram \underline{T}ranslations from Big Code}

\author{Binger Chen}
\affiliation{%
  \institution{TU Berlin}
  \city{Berlin}
  \country{Germany}
  }
\email{chen@tu-berlin.de}

\author{Ziawasch Abedjan}
\affiliation{%
  \institution{Leibniz Universit\"{a}t Hannover \& L3S Research Center}
  \city{Hannover}
  \country{Germany}
  }
\email{abedjan@dbs.uni-hannover.de}

\begin{abstract}

Program translation is a growing demand in software engineering.
Manual program translation requires programming expertise in source and target language. One way to automate this process is to make use of the big data of programs, i.e., Big Code. 
In particular, one can search for program translations in Big Code.
However, existing code retrieval techniques are not designed for cross-language code retrieval. Other data-driven approaches require human efforts in constructing cross-language parallel datasets to train translation models. 
In this paper, we present \system, a novel code translation retrieval system. 
We propose a lightweight but informative program representation, which can be generalized to all imperative PLs. Furthermore, we present our index structure and hierarchical filtering mechanism for efficient code retrieval from a Big Code database. 
\end{abstract}

\maketitle

\section{Introduction}\label{sec:intro}

Nowadays, numerous programs are being developed that require translations in other programming languages (PLs) to be further studied, reproduced, or applied on heterogeneous platforms. When the developers do not make the program translation efforts themselves, users have to manually rewrite the software in the needed PL, which is a time-consuming and error-prone process. 
Since traditional methods based on rule-based compilers or cross-language interpreters are hard-wired and require heavy human intervention for adaptation, the data-driven techniques are getting more traction.
Reuse of code from existing ``Big Code'' repositories, such as GitHub and Bitbucket, has the potential to support many programming tasks including program translation. 


Existing data-driven techniques for program translation are based on statistical models, such as 1pSMT~\cite{nguyen2013lexical}, mppSMT~\cite{nguyen2015divide} and Tree2tree~\cite{chen2018tree}, which train a program translation model. These approaches usually require a \emph{parallel dataset}, in which programs in different PLs are semantically aligned via manual efforts, to supervised learn the translation model. 
To avoid generating parallel datasets, recent work leverages a transfer learning approach from NLP~\cite{marie2020un}. They first train a model that denoises a randomly corrupted program and use it as a pre-trained program translation model, which is then optimized by back-translation method. However, by only relying on NLP features their approach neglects the special features of PLs. 
Furthermore, programs translated through the aforementioned approaches suffer from grammar mistakes because they are machine-generated programs. Therefore, they usually require additional human-supervision. 
Also, they are often confined to a few PLs because it is not trivial to extract general features that apply to every PL. 
Another promising direction is to retrieve similar code in target language directly from Big Code as potential translations. However, existing retrieval systems lack the proper capabilities for cross-language code retrieval~\cite{lv2015codehow,gu2018deep}. And instead of raw program input, they rely on queries  consisting of semantically expressive keywords, descriptions, or user specifications.


\indent In this paper, we propose \system, a novel \textbf{program translation retrieval system}. Given a raw program in a given source PL and a  target PL, \system efficiently retrieves similar programs as potential translations from Big Code, using \textbf{a generalizable program representation, an appropriate index structure, and a hierarchical filtering mechanism}.
Our approach does not require training data but can compete with existing translation models. 

\begin{figure}[htp]
  \centering
  \includegraphics[width=8.8cm]{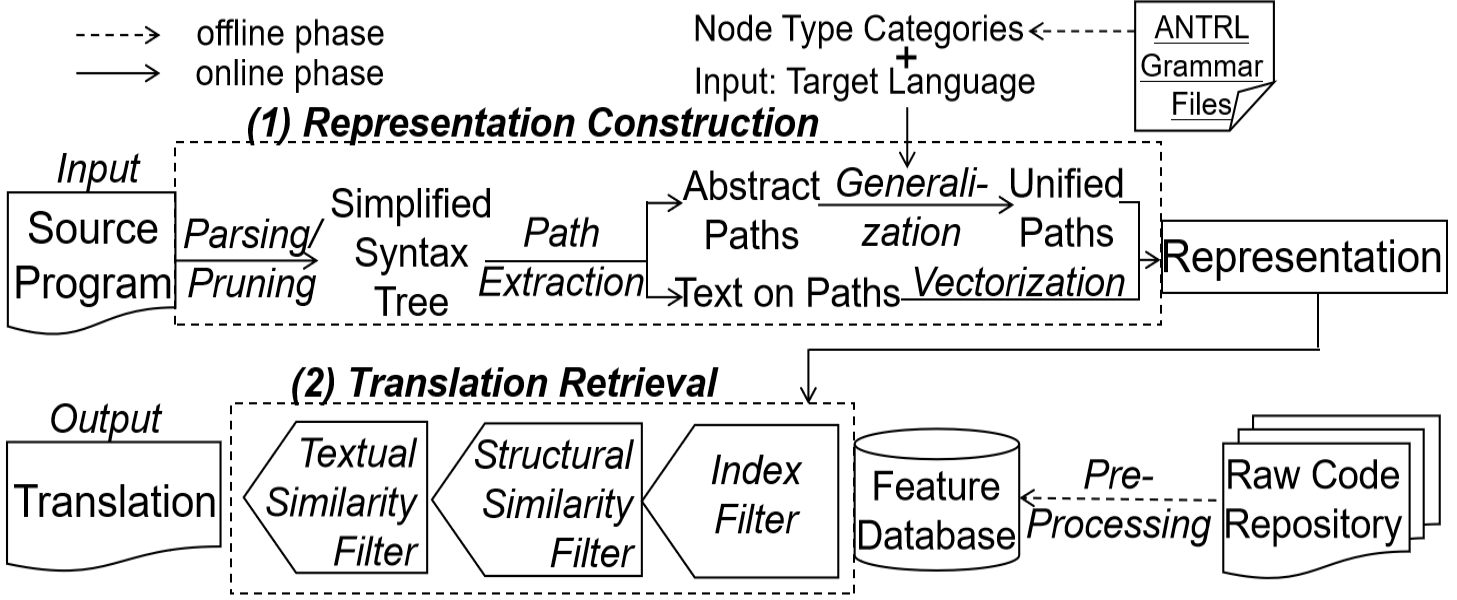}
  \caption{\system overview}\label{fig:system}
\end{figure}
\begin{figure}[!t]
  \centering
  \includegraphics[width=7cm]{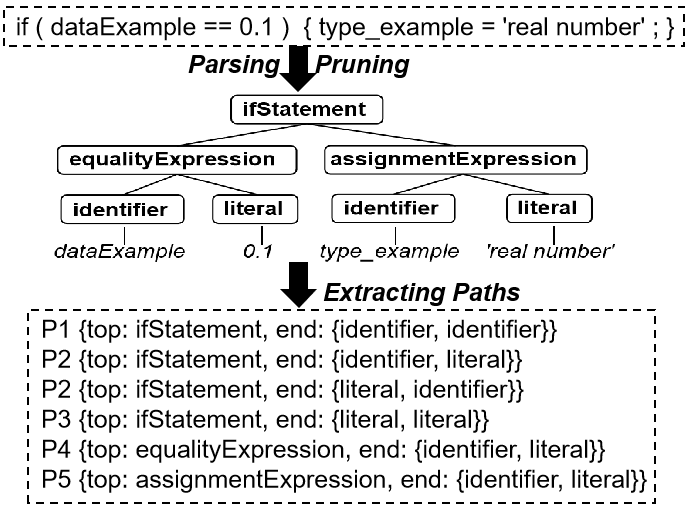}
  \caption{Simplified syntax tree \& Paths}\label{fig:repr}
\end{figure}
\section{Our approach}
We first discuss the necessary program representation and then the employed retrieval process of \system.
\subsection{Program Representation}
To identify cross-language code similarity, we need a unified representation that can be efficiently extracted from any given piece of code. 
Different from existing methods, \system considers both structural and textual features and their dependencies. 
\textbf{Structural features} can be captured by either a comprehensive concrete syntax tree~(CST) or an abstract syntax tree~(AST). The CST retains all the details, making it complicated and verbose with redundant information. The AST has more abstract but less informative syntax. And the abstraction strongly differs for different PLs. 
Thus, we fall back on the low-level CST as a basis and take the philosophy of AST to construct a unified abstract representation.
As shown in Figure~\ref{fig:system}, \system first employs a left-to-right parser to parse the source code and generate the original CST, which contains all the nodes and branches of the program structure. 
Then it is pruned to a simplified syntax tree to reduce the computation complexity. The simplified tree of a JavaScript code example is shown in Figure~\ref{fig:repr}.
The 2-D tree structure is further simplified to a set of 1-D paths that connect the elements on the tree: \system extracts \textit{abstract} paths by dropping all the intermediate nodes as the leaf-nodes and root node enclose the most critical information. The root node summarizes the whole path and the leaf-nodes directly indicate the content on this path. As the writing habit of programmers and the coding conventions for a PL might differ, \system ignores the order of left and right leaf-nodes.  Figure~\ref{fig:repr} shows the five extracted path types. 
Finally, \system generalizes these path types by matching all the node types across languages and substituting them with their category labels. 
Our approach leads to a more concise representation than CST but retains the original structural framework that is dropped in AST. 
Further,
\system only processes text that appears in the extracted paths. 
\system does not remove and tokenize numeric values, such as hard-coded floating points and integers, as they might be integral to the program. \system vectorizes these generated tokens through embeddings. 
The final \textbf{comprehensive representation} consists of three components: the list of path types that appear in a program, the frequency of different path types in a program, and the structure-dependent textual features based on the information of the relative position of text and structure. 

\subsection{Translation Retrieval}\label{sec:tr}
To make our approach scalable on big code, we implement a \textbf{hierarchical filtering mechanism} and a novel \textbf{index structure} for effective and efficient retrieval. 
The representation of each program is constructed offline and stored in a feature database. Our index structure is customized based on our representation.
Two similar programs usually share some common path types with similar frequencies. 
Thus, the index harbors the frequencies of each path type per program.  
As the frequencies may not be exactly the same, 
we divide the frequency into multiple buckets and use the bucket intervals as indexes. 
Because the frequency of each path in each program roughly obeys exponential distribution, we uniformly fill out fixed size buckets with different frequency intervals.
We name this as \emph{\textbf{p}ath-type-\textbf{b}ucket-\textbf{i}ndex (\pbi)}.
After using the index, \system filters candidates based on structural similarity first as it is more discriminative than textual similarity for a program and its translation. This can also facilitate the dependency between features to influence the subsequent textual similarity filter.
Further, \system runs textual similarity filter to determine the final candidate. For the source program and each candidate, \system calculates the weighted sum of both similarities.  
\begin{table}[!t]\small
\renewcommand\arraystretch{0.8}
  \caption{Program accuracy \& time cost per translation}
  \label{tbl:accuracy}
   \centering
    \begin{tabular}{ccccccc}
    \toprule
    \multirow{2}{*}{\textbf{Project}} & \multirow{2}{*}{\textbf{\system}} & \multicolumn{4}{c}{\textbf{Baselines}}\\
    \cmidrule(r){3-6} 
    & &\textsc{TransCoder} & \textsc{Tree2tree} &\textsc{mppSMT} &\textsc{1pSMT}  \\
    \midrule
    Lucene & 68.8\% &53.0\% & \textbf{72.8\%} & 40.0\% & 21.6\%  \\
    POI & 70.0\%&51.0\% & \textbf{72.2\%} & 48.2\% & 34.6\%  \\
    IText & \textbf{73.3\%}&45.9\% & 67.5\% & 40.6\% & 24.4\%  \\
    JGit & \textbf{74.5\%}&49.4\% & 68.7\% & 48.5\% & 23.0\%  \\
    JTS & \textbf{69.1\%}&43.2\% & 68.2\% & 26.3\% & 18.5\%  \\
    ANTLR & \textbf{71.9\%} &54.9\%& 58.3\% & 49.1\% & 11.5\%  \\
    \midrule
    \textbf{Time cost}&\textbf{0.20s} &0.36s  & 0.23s & 0.24s & 0.34s\\
  \bottomrule
\end{tabular}
\end{table}
\section{Experiments}\label{sec:expr}

\stitle{Experiments.} We apply our approach on a Java to C\# parallel dataset used in previous work~\cite{nguyen2013lexical,nguyen2015divide,chen2018tree}.  
We compare the results of effectiveness and efficiency of \system with state-of-the-art baselines 1pSMT, mppSMT, Tree2tree ,and TransCoder.
Our metric is \emph{program accuracy}~\cite{chen2018tree}: the percentage of the retrieved translations that are functionality the same as the ground truth in the dataset. 
The results in Table~\ref{tbl:accuracy} show that \system is competitive to all baselines despite the fact that \system is fully unsupervised and does not reuse existing data without training any models. Moreover, we observe that for the failed cases the translations tend to appear in the retrieved top 10 list.
Further, the efficiency of our retrieval based system is shown to be comparable to other baselines. We also compare our index \pbi to a simple path type index. \pbi leads to a runtime-improvement by two orders of magnitude at a scale of 3.8GB database.

\section{Conclusion and future work}\label{sec:cls}

We proposed \system, a code-retrieval approach to support program translation with Big Code, which is competitive with existing translation methods. We published our code on \url{https://github.com/BigDaMa/RPT}.

In future work, will augment the retrieval system with program generation to overcome the limitations of the database.

\section*{Acknowledgment}
This work was funded by the German Ministry for Education and Research as BIFOLD - Berlin Institute for the Foundations of Learning and Data (ref. 01IS18025A and ref. 01IS18037A).
\bibliographystyle{ACM-Reference-Format}
\bibliography{abbreviation}


\providecommand{\noopsort}[1]{}
\begin{thebibliography}{6}


\ifx \showCODEN    \undefined \def \showCODEN     #1{\unskip}     \fi
\ifx \showDOI      \undefined \def \showDOI       #1{#1}\fi
\ifx \showISBNx    \undefined \def \showISBNx     #1{\unskip}     \fi
\ifx \showISBNxiii \undefined \def \showISBNxiii  #1{\unskip}     \fi
\ifx \showISSN     \undefined \def \showISSN      #1{\unskip}     \fi
\ifx \showLCCN     \undefined \def \showLCCN      #1{\unskip}     \fi
\ifx \shownote     \undefined \def \shownote      #1{#1}          \fi
\ifx \showarticletitle \undefined \def \showarticletitle #1{#1}   \fi
\ifx \showURL      \undefined \def \showURL       {\relax}        \fi
\providecommand\bibfield[2]{#2}
\providecommand\bibinfo[2]{#2}
\providecommand\natexlab[1]{#1}
\providecommand\showeprint[2][]{arXiv:#2}

\bibitem[\protect\citeauthoryear{Chen, Liu, and Song}{Chen
  et~al\mbox{.}}{2018}]%
        {chen2018tree}
\bibfield{author}{\bibinfo{person}{Xinyun Chen}, \bibinfo{person}{Chang Liu},
  {and} \bibinfo{person}{Dawn Song}.} \bibinfo{year}{2018}\natexlab{}.
\newblock \showarticletitle{Tree-to-tree neural networks for program
  translation}. In \bibinfo{booktitle}{\emph{NeurIPS}}.
  \bibinfo{pages}{2547--2557}.
\newblock


\bibitem[\protect\citeauthoryear{Gu, Zhang, and Kim}{Gu et~al\mbox{.}}{2018}]%
        {gu2018deep}
\bibfield{author}{\bibinfo{person}{Xiaodong Gu}, \bibinfo{person}{Hongyu
  Zhang}, {and} \bibinfo{person}{Sunghun Kim}.}
  \bibinfo{year}{2018}\natexlab{}.
\newblock \showarticletitle{Deep code search}. In
  \bibinfo{booktitle}{\emph{ICSE}}. \bibinfo{pages}{933--944}.
\newblock


\bibitem[\protect\citeauthoryear{Lv, Zhang, Lou, Wang, Zhang, and Zhao}{Lv
  et~al\mbox{.}}{2015}]%
        {lv2015codehow}
\bibfield{author}{\bibinfo{person}{Fei Lv}, \bibinfo{person}{Hongyu Zhang},
  \bibinfo{person}{Jian-guang Lou}, \bibinfo{person}{Shaowei Wang},
  \bibinfo{person}{Dongmei Zhang}, {and} \bibinfo{person}{Jianjun Zhao}.}
  \bibinfo{year}{2015}\natexlab{}.
\newblock \showarticletitle{Codehow: Effective code search based on api
  understanding and extended boolean model}. In
  \bibinfo{booktitle}{\emph{ASE}}.
\newblock


\bibitem[\protect\citeauthoryear{Nguyen, Nguyen, and Nguyen}{Nguyen
  et~al\mbox{.}}{2013}]%
        {nguyen2013lexical}
\bibfield{author}{\bibinfo{person}{Anh~Tuan Nguyen},
  \bibinfo{person}{Tung~Thanh Nguyen}, {and} \bibinfo{person}{Tien~N Nguyen}.}
  \bibinfo{year}{2013}\natexlab{}.
\newblock \showarticletitle{Lexical statistical machine translation for
  language migration}. In \bibinfo{booktitle}{\emph{ESEC/FSE}}.
  \bibinfo{pages}{651--654}.
\newblock


\bibitem[\protect\citeauthoryear{Nguyen, Nguyen, and Nguyen}{Nguyen
  et~al\mbox{.}}{2015}]%
        {nguyen2015divide}
\bibfield{author}{\bibinfo{person}{Anh~Tuan Nguyen},
  \bibinfo{person}{Tung~Thanh Nguyen}, {and} \bibinfo{person}{Tien~N Nguyen}.}
  \bibinfo{year}{2015}\natexlab{}.
\newblock \showarticletitle{Divide-and-conquer approach for multi-phase
  statistical migration for source code (t)}. In
  \bibinfo{booktitle}{\emph{ASE}}. \bibinfo{pages}{585--596}.
\newblock


\bibitem[\protect\citeauthoryear{Rozi{\`{e}}re, Lachaux, Chanussot, and
  Lample}{Rozi{\`{e}}re et~al\mbox{.}}{2020}]%
        {marie2020un}
\bibfield{author}{\bibinfo{person}{Baptiste Rozi{\`{e}}re},
  \bibinfo{person}{Marie{-}Anne Lachaux}, \bibinfo{person}{Lowik Chanussot},
  {and} \bibinfo{person}{Guillaume Lample}.} \bibinfo{year}{2020}\natexlab{}.
\newblock \showarticletitle{Unsupervised Translation of Programming Languages}.
  In \bibinfo{booktitle}{\emph{NeurIPS}}.
\newblock


\end{thebibliography}

\end{document}